\documentclass[conference]{IEEEtran}
\IEEEoverridecommandlockouts
\usepackage{cite}
\usepackage{amsmath,amssymb,amsfonts}
\usepackage{algorithmic}
\usepackage{graphicx}
\usepackage{textcomp}
\usepackage{xcolor}
\usepackage{cite}
\usepackage{amsmath,amssymb,amsfonts}
\usepackage{algorithmic}
\usepackage{graphicx}
\usepackage{textcomp}
\usepackage{xcolor}
\usepackage{lipsum}
\usepackage{booktabs}
\usepackage{longtable}
\usepackage{tabularx}
\usepackage{array}

\usepackage{tcolorbox}
\usepackage{tabularx}
\usepackage{booktabs}
\usepackage{tcolorbox}

\usepackage{fancyhdr}
\begin{document}

\title{AI Mediators Regulate Emotion and Create Value in Disputes
}
\author{\IEEEauthorblockN{James Hale}
\IEEEauthorblockA{
\textit{University of Southern California}\\
Los Angeles, CA \\
jahale@usc.edu}
\and
\IEEEauthorblockN{Jonathan Gratch}
\IEEEauthorblockA{
\textit{Vanderbilt University}\\
Nashville, TN \\
jonathan.gratch@vanderbilt.edu}}

\maketitle
\thispagestyle{fancy}

\maketitle

\begin{abstract}
In conflict and disputes, especially, emotion acts as a salient force in influencing outcomes. Prior work shows negative affect can obstruct collaborative behaviors, which typically lead to ``win-win'' outcomes.  Thus, some suggest mediators may help regulate emotion and achieve joint gains.
With the proliferation of AI, we posit LLMs may perform well at this task, with the added benefit of better accessibility compared with a human mediator.
To examine the effectiveness of AI versus novice human mediators, we conduct a between-subjects experiment, where participants engage in a dispute mediated by a human, AI, or no mediator. We first analyze how well the mediators regulate emotions within a dispute -- finding AI mediators perform significantly better than humans at reducing negative emotion. We next examine whether AI mediators facilitate disputants better realizing joint gains in disputes with high integrative potential (IP) -- we find a marginally significant interaction between IP and condition (AI versus human), indicating LLMs may outperform humans at aiding disputants realize joint gains. Lastly, we perform an analysis of the messages the mediators sent, finding the AI sent significantly more messages suggesting trade-offs compared to the humans.

\end{abstract}


\section{Introduction}
Disputes, a form of negotiation, take a mixed-motive structure -- i.e., each side has competing goals, as well as shared interests -- where one can only maximize their self-interest through compromise \cite{may2018compromise}. However, literature characterizes disputes as a more intense and emotional form of negotiation, where conflict centers around a perceived wrong -- e.g., one side raises a claim, which the other side rejects \cite{felstiner2017emergence}, causing conflict in an existing relationship, disagreement on facts, and a lower chance at resolution. The intensity of the conflict can lead disputants to respond to anger with anger \cite{pruitt2007conflict}, and become entrenched in their positions -- sometimes leading to physical threats~\cite{brett1998breaking, halperin2008group,pruitt2007conflict}. This contrasts with deal-making negotiations, which researchers more often study, where displays of anger can induce concession~\cite{van2004interpersonal,de2011effect}. These innate properties of disputes make compromise and resolution uniquely challenging.

There exists ample evidence that affect influences value creation in negotiation -- and by extension, disputes -- where negative affect encumbers joint gains \cite{carnevale1986influence, isen2015relationship}. 
In negotiation, dyads can achieve joint gains through collaborative behavior such as information exchange~\cite{thompson1991information}, log rolling~\cite{tajima2001logrolling}, etc. However, the emotionally charged nature of disputes makes collaborative behavior less likely. 
For example, Carnevale and Isen found that positive affect in negotiation decreased ``contentious tactics'' between parties and improved joint outcomes \cite{carnevale1986influence}. Further, Isen \cite{isen2015relationship} finds that increasing positive affect facilitates more creative problem-solving across myriad domains, including negotiation. 
Thus, achieving joint gains requires careful management of emotional dynamics.

In light of this, many disputes engage mediators to facilitate resolution and achieve ``win-win'' outcomes\cite{bingham2004employment,jacobs2002mediators}.
There exists a diverse array of guidelines as to the purpose of a mediator, and how they should operate \cite{riskin1994mediator,izumi2010implicit,nolan2013judicial}. A common perspective, for example, emphasizes the importance of neutrality in mediation, and that mediators should not impose a solution on parties -- rather, they should help parties resolve the dispute voluntarily, preserving their self-determination \cite{izumi2010implicit,nolan2013judicial}.   
Further, there exists some work on the importance of allowing and effectively managing emotion expression during mediated disputes\cite{douglas2014attitude,picard2013exploring,hoffman2011mediation}.
If a mediator can competently regulate emotions, they may induce parties to leverage more collaborative behaviors, which can lead to better joint outcomes\cite{greco2024naming}. 
A mediator, to find a ``win-win'' solution, may seek to prevent escalation of conflict\cite{pruitt2007conflict}, clear any miscommunication between parties\cite{van2010dealing}, and move a disputant away from unreasonable demands\cite{kirgis2014bargaining}. We focus on these goals in our work -- prompting human and AI mediators to consider these reasons when attempting to facilitate a resolution.

Our work makes several contributions as it relates to affective computing. We examine the potential of LLMs to operate as mediators in contentious and emotional disputes,  evaluating their ability to influence emotional dynamics -- i.e., increasing positive affect and decreasing negative emotions. As we previously made a connection between affect and inducing collaboration, we examine whether we see that reflected in our sample -- i.e., we test whether AI mediators better regulate emotion \textit{and} if we see superior realization of joint gains.

\begin{figure*}[t]
\centering
\begin{tcolorbox}[
    colback=gray!5,
    colframe=black!20,
    boxrule=0.5pt,
    arc=2pt,
    left=6pt,
    right=6pt,
    top=6pt,
    bottom=6pt,
    width=\textwidth
]
\footnotesize
\begin{tabularx}{\textwidth}{@{}>{\bfseries}l@{\hspace{0.75em}}X@{}}
Buyer: 
& I can't remove my review without compensation. Offer a partial refund, remove your false review, and apologize, and we can move forward. \\[0.4em]

Mediator: 
& Seller, maybe you would consider a refund if Buyer returned the item? \\[0.4em]

Seller: 
& I would absolutely give a full refund if the buyer returned the item. My issue is that this buyer didn't even contact us to let us know there was an issue; we could have already resolved this with a return/refund, even though it is fully their fault that they didn't read my description. My issue is that you are submitting dishonest reviews and you want me to compensate you for that. I'd be happy to do a full refund upon receiving my item back, and I want this review taken down and an apology. \\[0.4em]

Buyer: 
& I'll return the item for a full refund, but the false review must be removed too. Apologizing isn't on the table unless you admit your mistake. \\[0.4em]

Seller: 
& Actually, you know what. I don't negotiate with customers like this. You should have read thoroughly, and when you chose not to, you should have reached out to me to discuss a return and a refund. You are making this more of a headache than it's worth. My loyal customers will see that you're just one of many who try to get over on small businesses. Take care. \\[0.4em]

System: 
& \emph{Buyer chose to walk away.}
\end{tabularx}
\end{tcolorbox}
\caption{Example snippet from a mediated dialogue we collected, where the buyer and seller could not agree, even with the help of a human mediator.}
\label{fig:dialogue-impasse-example}
\end{figure*}

\section{Related Work}

\subsection{Affect in Dispute Resolution}
The conflict literature underscores the salience of affect in influencing whether parties can reach an agreement~\cite{pruitt2007conflict,
brett2007sticks,
morris2000emotions, rakshit2025emotionally}.
Morris and Keltner \cite{morris2000emotions} propose that emotions in dyadic settings, such as negotiations of disputes, serve to communicate one's internal state to solve problems arising within the dyad -- this, they argue, contrasts with the perspective that emotion \textit{just} impacts one's decision-making abilities.
Further, Pruitt ~\cite{pruitt2007conflict} proposes the notion of escalatory spirals in disputes, where one party expresses anger, the other responds in kind -- this emotional spiral may serve to derail any chance of resolution as disputants become entrenched in their positions. 
However, Friedman et al. \cite{friedman2004positive} found that the effect of expressions of anger in mediated online disputes depended heavily on one's vulnerability -- e.g., while generally one responds to anger with anger, a vulnerable disputant may not respond to anger in the same way. 
Additionally, Brett et al. \cite{brett2007sticks}, in an analysis of online dispute dialogues, found a lower likelihood of resolution in disputes where one attacks another's face with negative emotions. 
Further, Rakshit et al. \cite{rakshit2025emotionally}, in a similar setting to what we use, demonstrate the power of automatically (LLM) recognized emotion in explaining the variance in a dispute's subjective outcomes \cite{curhan2006people} -- i.e., how disputants felt about the relationship with their partner, their performance, the outcome, and the fairness of the process. 

Thus, a mediator trying to facilitate a resolution in a dispute should understand the salience of emotional dynamics in disputes, not only as they relate to objective and subjective outcomes, but also in how they serve to communicate with one's partner.
Consequently, there exists literature on how professional mediators in disputes manage emotion in practice \cite{douglas2014attitude, greco2024naming,picard2013exploring,hoffman2011mediation}. 
Douglas and Coburn \cite{douglas2014attitude} surveyed professional mediators regarding their strategies for managing emotion in disputes and found that the majority of mediators promote free expression of emotion. 
Picard and Siltanen emphasize, ``Emotion is the pathway to discovering the values that are threatened through the conflict dynamic—values that hold the key to achieving successful and lasting change'' \cite{picard2013exploring}.


\subsection{Integrative Potential in Disputes}
There exists prior work on the influence of positive and negative affect on collaborative behavior and joint gains\cite{carnevale1986influence, isen2015relationship}. For example, Carnevale and Isen \cite{carnevale1986influence} found that positive affect in negotiation increased collaborative behaviors, such as creative problem solving, which yielded improved joint gains -- Isen \cite{isen2015relationship} similarly finds positive affect improved creative problem solving. 
Further, Allred et al. \cite{allred1997influence} found that negotiators who felt more negative and less positive emotions for each other earned fewer joint gains. 
Thus, there exists a potential pathway for mediators to manage emotions, thus inducing collaborative behaviors to improve mutual benefits.

\subsection{AI Mediation}
The importance of mediators in disputes combined with the proliferation of AI has caused the research community to examine the extent to which LLMs can act as mediators. 
For example, Hale et al. \cite{hale2025ai} task an LLM to run over pre-collected dispute dialogues, asking what it \textit{would} have done if it were a mediator -- they found the LLM attended to the disputes' self-reported emotion, intervened more so in disputes ultimately ending in impasse, and that third party raters preferred the LLM-generated mediated messages compared to novice human ones. 
In a follow-up study, they found, in a three-party interaction (two human disputants, and a human mediator) that human mediators whose behavior agreed more with the LLMs' suggested behavior afterwards, achieved better outcomes \cite{hale2026can}.
Similarly, Tan et al. \cite{tan2024robots} found the LLM-driven mediators outperform humans in selecting appropriate intervention strategies and in crafting compelling messages in a curated dispute dataset.
However, to our knowledge, little work exists on an LLM mediator's ability to regulate emotion and facilitate joint gains in heated disputes compared against novice humans, and no mediation baselines.

\section{Research Questions}
Thus, we posit these research questions:
\begin{itemize}
\item \textbf{RQ1:} Can LLM mediators, relative to novice humans, more effectively regulate emotion, and dampen escalatory spirals in disputes?
\item \textbf{RQ2:} Can LLM mediators effectively facilitate trade-offs in disputes of varying potential for joint gains?
\end{itemize}

\section{Methodology}
\subsection{Dispute Task}
We use the dispute scenario proposed in Hale et al.'s \cite{hale2025kodis} \texttt{KODIS} corpus, an open-source collection of simulated dispute dialogues leveraging a buyer-seller dispute. The authors validate the effectiveness of their scenario via the presence of quintessential dispute markers -- namely, a high impasse rate (approximately one in five), and the appearance of various emotional dynamics \cite{rakshit2025emotionally} -- e.g., escalatory spirals.

\subsubsection{Disputant Perspective} 
Specifically, \texttt{KODIS} prompts participants to role-play as disputing buyers or sellers, where the buyer initially purchases a basketball jersey from an online platform for their sick nephew. The buyer believes they received the incorrect jersey in the mail, and requests a refund from the seller; the seller denies their refund request, stating the buyer misread the original listing. Their relationship further deteriorates as each side levies negative reviews against the other. See Hale et al.'s work \cite{hale2025kodis} for details on the instructions the seller role-player would read before entering the dispute.








We tasked disputants to resolve a dispute over these issues: 
\begin{itemize}
    \item \textbf{Refund:} The buyer wishes to receive a refund, while the seller wishes to avoid granting one. This issue has three levels -- \textit{no refund}, \textit{partial refund}, and \textit{full refund}.
    \item \textbf{Buyer's Review:} The buyer posts a negative review of the seller. The seller wishes for the buyer to remove the review, while the buyer would like to keep it up. This issue has two levels -- \textit{keep}, and \textit{retract}.
    \item \textbf{Seller's Review:} Similar to the previous issue, this concerns the review the seller posted. Here, the seller would like to keep this review, while the buyer would like it taken down. This issue has the same two levels.
    \item \textbf{Apologies:} Lastly, the two parties can negotiate over whether they issue formal apologies to each other. This issue has four levels -- \textit{buyer apologizes}, \textit{seller apologizes}, \textit{buyer does not apologize}, and \textit{seller does not apologize}. 
\end{itemize}
Before the dispute, we ask participants to select which issues they prefer by allocating 100 points among them. This contrasts with other studies which assign payoff functions to participants, and allows for disputes of varying integrative potential. We scored participants based on their stated preferences, and whether they achieved those objectives (see Section~\ref{sec:participants} for details on payment and Section~\ref{sec:scoring} for scoring details). 
To incentivize fidelity to stated preferences, we reward participants with a monetary bonus contingent on achieving objectives self-defined as more important. After the task, we ask participants to restate their preferences --
Spearman rank correlation yielded a moderate ($\rho = .58$) average correlation between before and after task preferences.


\begin{figure}[t]
\centering
\footnotesize
\begin{minipage}{0.95\linewidth}
\begin{tcolorbox}[
    colback=gray!5,
    colframe=black!20,
    title=Mediator Instructions,
    fonttitle=\bfseries,
    boxrule=0.5pt,
    arc=2mm,
    left=2mm,
    right=2mm,
    top=2mm,
    bottom=2mm
]
You will play the role of a mediator in a buyer/seller purchase dispute. Your goal is to allow participants to resolve their dispute on their own if possible, but to intervene if necessary. 
\vspace{.25cm}

Some reasons to intervene include:

\begin{itemize}
    \item \textbf{Escalation of Conflict:} If the conversation becomes heated with parties resorting to personal attacks or hostile language
    \item \textbf{Impasse:} When parties reach a deadlock and are unable to move forward.
    \item \textbf{Miscommunication:} If there are signs that the parties are misunderstanding each other’s points.
    \item \textbf{Unreasonable demands:} If one party is making unreasonable demands that the other party can’t possibly meet.
    \item \textbf{Invocation:} If one party asks the mediator to interject.
\end{itemize}

You do not need to intervene every turn, and should consider how recently you've intervened before making a decision.
\end{tcolorbox}
\end{minipage}
\caption{Prompt read by mediator before the task.}
\label{fig:mediator-role-play-instructions}
\end{figure}

\subsubsection{Mediator Perspective}
Whereas the original \texttt{KODIS} corpus did not include mediators, we craft instructions for the humans and LLM taking that role. Figure~\ref{fig:mediator-role-play-instructions} outlines the instructions the human mediator read before entering the dispute, and the LLM's prompt contained the same text.
Participants re-enter their preferences when moving from the survey to the online platform, which the human and LLM mediators can access when evaluating a dispute and deciding whether to intervene and what to say. 
Mediators chose to intervene or not after each utterance from a disputant, and we task them not only with aiding each side in reaching agreement, but also in helping facilitate agreements beneficial to both sides.


\subsection{Experiment Design}
\subsubsection{Participants}\label{sec:participants}
Our university's Institutional Review Board approved
the experimental design and classified it as minimal risk.
We leveraged a crowd-sourcing platform, Prolific, to recruit participants for our study; and recruited participants ($N=164$ as seller disputants, and $N=50$ as mediators) with the United States as their country of residence. Our sample yielded a gender breakdown of 55\% male, 43\% female, 1\% other; and an ethnic breakdown of 60\% White, 16\% Black, 11\% Asian, 5\% Mixed, 5\% other, and 1\% who preferred not to say\footnote{For 2\% of participants, their ethnic data from Prolific expired.}. Lastly, participants had an average age of 37.9 years with a standard deviation of 12.6.

Participants received a base payment of \$3.50. Disputants had an opportunity to earn an extra \$3.00 if they achieved the most beneficial deal; mediators, in cases of agreement, could earn between \$1.50 and \$3.00 extra depending on the proportion of integrative potential realized. If the disputants walked away, neither they nor the mediator received a bonus.

\subsubsection{Pre-Task}
Participants began the study by consenting to participate, and answering some demographic questions. Next, they read a role-play prompt, as previously outlined; depending on their randomly assigned role, they either read the disputant or mediator instructions. For simplicity, we always assign an LLM (\texttt{GPT4o}) to play the buyer side, prompting it with similar instructions to what the seller read, and instructing it to speak conversationally -- thus, a human participant always plays the seller. 
From \texttt{KODIS} \cite{hale2025kodis}, impasses typically occur when a seller responds to a buyer’s expression of anger with anger; however, many buyers don’t show anger. Thus, to induce situations requiring mediation, we fix an AI confederate to act as the buyer, common in negotiation research \cite{van2004interpersonal, de2011effect}. 

Inputting one's preferences over the issues (\textit{refund}, \textit{buyer's review}, \textit{seller's review}, and \textit{apologies}) remains the final step before entering the dispute, for those playing as the seller\footnote{For the buyer side, played by the AI, we use the following preferences, which we derive from the mean preferences in Hale et al.'s \texttt{KODIS} \cite{hale2025kodis} -- 57 points of 100 for receiving a refund, 16 points for having the seller retract their review, 14 points for keeping their review up, and 13 points for receiving a formal apology. Humans role-playing sellers most preferred avoiding giving a refund ($M=38.9$, $SD=20.4$), followed by getting the negative review against them removed ($M=30.3$, $SD=17.1$), followed by keeping their review against the buyer ($M=15.7$, $SD=10.4$), and finally denoting receiving a formal apology ($M=15.1$, $SD=12.8$) as least important.} -- this step allows disputes with varying degrees of integrative potential. The mediator skips this step and proceeds to the dispute. Participants then proceed to the online platform, 
and wait in a lobby to match with a partner -- i.e., mediators enter the lobby to match with another human seller, and sellers enter to match with a human mediator or an AI mediator depending on condition\footnote{In the case of a seller matching with an AI mediator -- i.e., the mediator and buyer are AI -- we keep them in the waiting room for a period of time sampled from a normal distribution centered at five minutes, and with a standard deviation of two minutes. In either case, participants wait in the matching lobby for a maximum of ten minutes before we fill in missing roles with LLM based agents.}. 
\subsubsection{Intervention}
To answer our research questions surrounding the effectiveness of AI mediators, our experiment takes a between-subjects design with three conditions, where human sellers enter a dispute against the buyer in one of the following mediation configurations. 
\begin{itemize}
    \item \textbf{No Mediation:} Disputants attempt to negotiate an agreement without a mediator (AI or human) monitoring the conversation. 
    \item \textbf{AI Mediation:} An LLM mediator monitors and can intervene in the dispute.
    \item \textbf{Human Mediation:} A human mediator, another crowd-sourced worker from Prolific, monitors the dispute and can interject.
\end{itemize}
Of note, the participants assigned to act as a mediator will always mediate disputes with a human as the seller -- unless the max time expires in the matching lobby, in which case they mediate two AI disputants, though we do not analyze those mediations here. This study intends to study the effect of different mediators on \textit{human} disputant outcomes. 

We leveraged \texttt{gpt-5-mini-2025-08-07} and \texttt{gpt-5-2025-08-07}\footnote{For these, and all other LLMs, we use default parameters, including a temperature of 1.} to act as the AI mediator, with the \texttt{mini} model monitoring and deciding whether to intervene and the larger model crafting the messages when triggered -- we did this to reduce latency in the model's response\footnote{The smaller model has lower latency on average, and the smaller one triggers the larger one. In this way, the larger model does not generate a message when not needed, passing to the next disputant quickly.}. We give the monitoring model a prompt nearly identical to Figure~\ref{fig:mediator-role-play-instructions} along with the dialogue history, instructing it to output a rating (one to ten) whether it should intervene, with any score over a seven\footnote{We derive this threshold from a prior offline mediation study on the \texttt{KODIS} \cite{hale2025kodis} corpus, where we experimented with various intervention thresholds -- finding a value of seven yielded more interventions in cases of impasse.} triggering the larger model to craft and send a message. We prompt the model to act ``neutral'' and to guide parties to a ``voluntary agreement.'' 

\subsubsection{Dispute Task}
Participants communicate using natural language on a chat interface. We impose strict turn-taking, where parties can only send one message at a time, and must wait for their counterpart to respond before sending another.
After eight total messages exchanged between the buyer and seller, either side could use the menu on the right side of the interface to propose a final deal; or, if they felt they could not resolve the dispute, they could walk away. The dispute ends when either one side proposes an agreement that the other accepts, or if one side chooses to walk away. 
Meanwhile, the mediator monitors the dispute's progress, seeing new messages as they arrive, and has the option of intervening with a message after each utterance sent by a disputant -- otherwise, they can decline and pass the turn to the next party. Notably, the mediator -- both AI and human -- has access to the preferences of each party, and may use that to facilitate value creation. 

\subsubsection{Post-Task}
After returning to the survey from the online dispute platform, participants answered a few questions regarding their impression of the system. 
Lastly, participants read a debrief, telling them the nature of their counterparts. If both other actors in the dispute were AI, they read a message that said, ``In this study, all other parties were AI-powered agents that generated messages automatically in real time.'' Otherwise, if an LLM only powered the buyer, they read, ``In this study, one of the negotiation participants you interacted with was an AI-powered agent that generated messages automatically in real time. The buyer was AI.''
\begin{table*}[tbh!]
\centering
\footnotesize
\setlength{\tabcolsep}{6pt}
\renewcommand{\arraystretch}{1.2}
\begin{tabularx}{\textwidth}{@{}lX@{}}
\toprule
\textbf{Mediator} & \textbf{Trade-off Suggestion} \\
\midrule
AI &
I’m hearing Buyer wants a full or partial refund and the seller to update/remove their negative review, while Seller wants an apology and no refund; would you both consider either a return for a full refund with Buyer covering return shipping or a partial refund with Buyer keeping the jersey and both of you updating your reviews to reflect the resolution? \par\vspace{0.6em}




To break the deadlock, would you accept either (A) Buyer posts a brief apology; Seller removes/updates the review the same day and provides a 20\% store credit, or (B) Buyer returns the jersey for a full refund with Buyer covering return shipping and both of you post neutral resolution updates? \\
\midrule
Human &
We have some areas of disagreement, but in order to reach some type of deal, both the buyer and seller will have to make some compromises. \par\vspace{0.6em}

If a refund of any sort is not possible, something else will have to be given up in order for some type of agreement to be reached. \par\vspace{0.6em}



\\
\bottomrule
\end{tabularx}
\caption{This table shows examples of AI and human mediators suggesting trade-offs in the dispute.}
\label{tab:tradeoff_examples}
\end{table*}
\subsection{Measures}

\subsubsection{Objective Measures}\label{sec:scoring}
We examine two measures as it relates to objective outcome -- impasse rate (i.e., the proportion of disputes ending in walk away), and points scored by each side given the outcome and their stated preferences (with walk-aways coded as zero). For points, take the following:
\begin{align}\label{fun:score}
    U_a = \sum_{i\in I} w_{i,a} * \ell_{i,a}
\end{align}
Where $a\in\{\text{buyer}, \text{ seller}\}$; $I$ is the set of all issues under dispute; $w_{i,a}$ is the value disputant $a$ ascribes to issue $i$ ($w_{i,a}\in[0,100]$); and $\ell_i,a$ is the level achieved on issue $i$ from $a$'s perspective, with zero as the worst outcome and one as the best\footnote{Most issues have two levels which map to zero or one, with the exception of \textit{refund}, which has three -- thus \textit{partial refund} maps to 0.5.}. 
Lastly, we also examine the realized joint gains yielded by the dispute's outcome -- joint points. We define that as the sum of points scored by the buyer and seller.

\subsubsection{Integrative Potential}
Integrative potential quantifies the extent to which parties can create joint gains in the dispute -- i.e., the extent to which parties can mutually obtain better outcomes via trade-offs. For example, if the buyer has a strong preference for receiving a refund, while the seller most wants the negative review removed, there exists opportunity for joint gains. While the buyer always uses the same preferences, the sellers possessing different preferences each round allows for varying levels of integrative potential. 
We calculate integrative potential (IP) by reverse scoring the cosine similarity of the preferences of each side; i.e., given the buyer's ($\Vec{X}_{buyer}$) and seller's preferences ($\Vec{X}_{seller}$): 
 \begin{align}
     \text{IP} = 1 - \dfrac{\Vec{X}_{buyer} \cdot \Vec{X}_{seller}}{\|\Vec{X}_{buyer} \| \| \Vec{X}_{seller}\|}
 \end{align}
This metric ranges from zero to one, where disputes with highest potential for value creation have $IP=1$.

\subsubsection{Affective Measures}
Following Rakshit et al.'s \cite{rakshit2025emotionally} approach, we use LLMs to annotate the emotional content of these disputes. Specifically, we use \texttt{GPT4.1} to annotate each line of dialogue for six emotion categories relevant to disputes -- \textit{anger}, \textit{joy}, \textit{fear}, \textit{sadness}, \textit{surprise}, and \textit{compassion}. 
We opt for this set as it includes emotions relevant to intense disputes -- e.g., \textit{compassion}, and \textit{anger}~\cite{Allred97compassion,TingToomey14-compassion, pruitt2007conflict}. 
Specifically, the model takes an utterance plus all previous messages, and ascribes a vector of emotion intensity values such that they sum to one -- there also exists a \textit{neutral} category for the model to use if warranted. For example, here are two examples of emotional messages sent by participants:
\begin{itemize}
    \item Seller: \textit{NO REFUND ALL SALES ARE FINAL PERIOD} $\rightarrow$ 0.9 \textit{anger}, 0.1 \textit{neutral}. 
    \item Seller: \textit{I stated I will remove my review and I apologize for any stress this has caused.} $\rightarrow$ 1.0 \textit{compassion}.
\end{itemize}

We run a small user-study to gauge the alignment of these emotion labels with human annotators. We randomly select 50 seller (human) utterances\footnote{We select utterances such that the LLM's evaluation of \textit{Neutral} is less than one -- i.e., it places weight on other emotions.}, and instruct Prolific workers ($N=255$)\footnote{We pay annotators \$1 for a task estimated to take less than five minutes.} to capture the extent to which the given utterance\footnote{We also give the dialogue history to that point.} expresses each emotion under consideration. Specifically, they allocate 100 points over the seven emotion labels. We ensure each selected utterance receives at least four annotations ($M=5.1$). 
In our sample, by Pearson Correlation, we see LLM annotations of several emotions relevant to disputes moderately correlate with the average human rating -- e.g., \textit{Anger} ($r = .58$, $p<.001$), and \textit{Compassion} ($r = .44$, $p=.001$);
\textit{Joy} ($r = .32$, $p=.02$), \textit{Sadness} ($r = .10$, $p=.47$), and \textit{Neutral} ($r = .39$, $p<.01$) exhibit weak positive correlation;
\textit{Fear} ($r =.03$, $p=.82$) does not yield positive correlation. 
\textit{Surprise} occurred rarely, and did not appear greater than zero in our sample; we also do not fold it into the \textit{positive} or \textit{negative} emotion groups -- thus we exclude from this analysis.
Overall, correlation differed by emotion, with \textit{anger} and \textit{compassion} exhibiting moderate agreement. 

\subsubsection{Dialogue Coding}\label{sec:coding}
We use LLMs to code the collected dialogues for speech acts -- specifically, quantifying the extent to which mediators suggest trade-offs, given the integrative nature of these disputes. J{\"a}ckel et al. propose NegotiAct \cite{jackel2024negotiact}, which we adapt to code mediator messages, and add a code getting at whether mediators send messages suggesting trade-offs. In this paper, we only consider that trade-off code -- we operationalize a dependent variable wherein we examine the proportion of opportunities the mediator had to intervene (i.e., the number of disputant utterances, as it could intervene after each) where it sent a message suggesting a trade-off.




\section{Results}

\subsection{Effect on Impasse}

We first examine whether there exist differences in the impasse rate by condition.
Table~\ref{tab:outcomes-by-condition} illustrates the counts of resolved disputes versus impasses -- we see the AI (7.1 interventions per dialogue) condition has the lowest impasse rate (47.3\%), relative to the no mediation (54.3\%) and human mediated (6.2 interventions per dialogue) conditions (58.0\%). However, a Chi-squared test did not yield a significant result for impasse rate ($\chi^2(2,N=164)=1.262$, $p=.53$). 

\begin{table}[tbh!]
\centering
\begin{tabular}{lrrrr}
\toprule
Condition & Resolution & Walk-away & Total & Impasse Rate \\
\midrule
AI Mediator & 29 & 26 & 55 & 47.3\% \\
No Mediation & 27 & 32 & 59 & 54.2\% \\
Human Mediator & 21 & 29 & 50 & 58.0\% \\
\bottomrule
\end{tabular}
\caption{Depicts the differences in impasse rate (percent of disputes ending in walk-away) by mediator condition. }
\label{tab:outcomes-by-condition}
\end{table}

\subsection{AI Mediators Better Control Emotion}\label{sec:emotion}
In this section, we analyze the impact of mediation on a human disputant's emotion. For a given dialogue, we consider the mean emotion intensity score of each emotion expressed by the seller (human-side) directly before and after each mediator message. For simplicity, we combine the emotions into \textit{negative} and \textit{positive}, as explained below\footnote{We omit \textit{surprise} from the \textit{negative} and \textit{positive} categories, as prior work does not clearly denote it as \textit{positive} or \textit{negative} \cite{noordewier2013valence}.}. We compare the AI-mediated condition against the human-mediated and no mediation conditions -- as the condition without mediation does not contain mediation messages to compare against, we use \texttt{gpt-5-mini-2025-08-07} to indicate where it \textit{would have} intervened, in the same manner as the AI during the live dispute sessions -- we intend for this to serve as a counterfactual baseline, to demonstrate the extent to which mediation-inducing emotion decreases without intervention. Thus, we remind the reader that this condition, unlike the other two, does not contain mediation messages. 
\begin{figure}
    \centering
    \includegraphics[width=\linewidth]{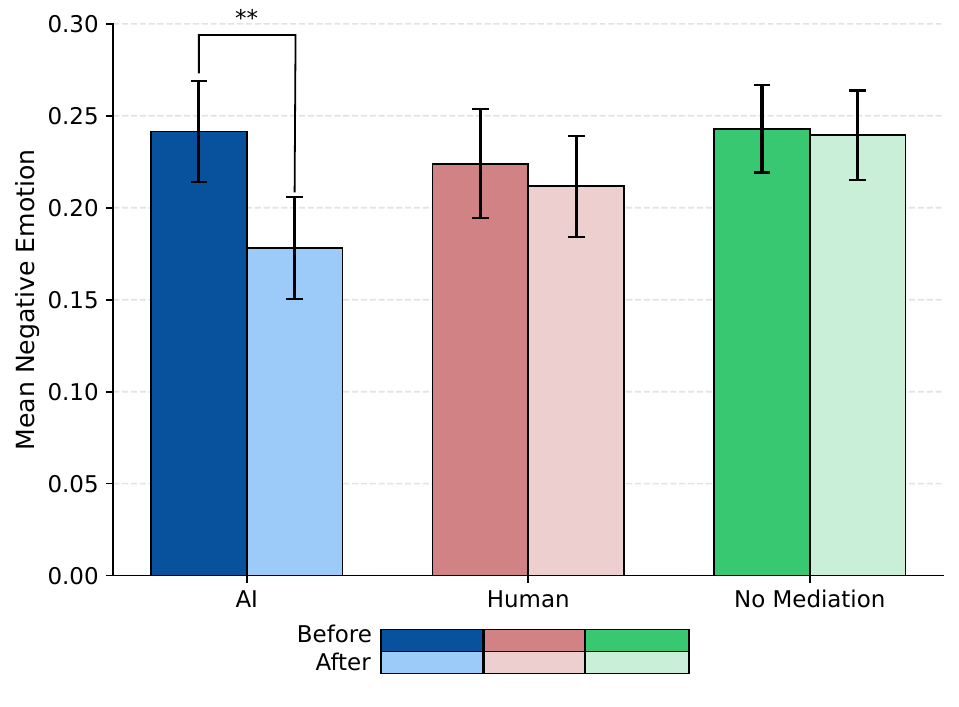}
    \caption{This depicts a significant interaction between condition (AI mediator versus human mediator versus no mediator), and time (before or after mediator message) for negative emotions.}
    \label{fig:emotion_interaction}
\end{figure}

\subsubsection{Dampening of Negative Emotions}
We first examine the impact of mediation on negative emotions -- specifically combining scores for \textit{anger}, \textit{fear}, and \textit{sadness} into a single measure of negative emotion. We next conduct a repeated measure analysis of variance (RM ANOVA) to test whether there exists an interaction between Mediation Type and Time (before / after). This test showed a significant main effect of time ($F(1,154)=9.04$, $p<.01$), indicating mediation worked to reduce negative emotion overall\footnote{Running the same RM ANOVA for each negative emotion, focusing on this interaction, we see \textit{anger} ($p=.07$) and \textit{fear} ($p=.07$) act as the strongest effects, compared to \textit{sadness} ($p=.56$). For each individual emotion, the seller's emotional intensity drops after the mediator's intervention.} -- before ($M=0.24$, $SD=0.19$) versus after ($M=0.21$, $SD=0.19$). It also yielded a significant ($F(2,154)=4.57$, $p=.01$) interaction between time and mediation condition. A post-hoc t-test revealed the AI mediator significantly ($t=3.20$, $p<.01$) reduced negative emotions after ($M=0.18$, $SD=0.20$) sending a message compared to before ($M=0.24$, $SD=0.20$), whereas the  human mediator and condition without mediation did not. Figure~\ref{fig:emotion_interaction} illustrates this interaction. 
Further, while ANOVA generally suffices for large samples, we also conduct non-parametric tests, replicating the interaction between Time and Mediator Type by Kruskal-Wallis ($H=6.57$, $p=0.038$); further, there remains a significant difference of Time for the AI mediator by Wilcoxon signed-rank ($p<.001$).

\begin{table*}[t]
\centering
\footnotesize
\setlength{\tabcolsep}{5pt}
\renewcommand{\arraystretch}{1.15}
\label{tab:mediator-code-proportions}

\begin{tabular}{lccccccccc}
\toprule
\textbf{Mediator} 
& \textbf{SACT} 
& \textbf{SMOD} 
& \textbf{PROC} 
& \textbf{TRDO} 
& \textbf{ASUB} 
& \textbf{APOS} 
& \textbf{CLAR} 
& \textbf{TIME} 
& \textbf{ALIS} \\
\midrule
AI 
& .474 (.036) 
& .409 (.038) 
& .415 (.041) 
& .383 (.036) 
& .289 (.040) 
& .228 (.030) 
& .195 (.023) 
& .239 (.029) 
& .113 (.015) \\
Human 
& .192 (.024) 
& .067 (.013) 
& .049 (.013) 
& .072 (.014) 
& .020 (.007) 
& .072 (.016) 
& .061 (.013) 
& .009 (.005) 
& .040 (.012) \\
\bottomrule
\end{tabular}

\vspace{0.5em}
\begin{minipage}{0.97\textwidth}
\footnotesize
\textit{Note.} SACT = suggesting action; SMOD = suggesting offer modification; PROC = procedural suggestion; TRDO = suggesting trade-offs; ASUB = asking for substantiation; APOS = asking for positional information; CLAR = clarification; TIME = time management; ALIS = active listening.
\end{minipage}
\caption{Mean proportion (with standard error of the mean) of opportunities each mediator took to send a message, fitting various codes from NegotiAct \cite{jackel2024negotiact}.}
\label{tab:negotiact}
\end{table*}

\begin{figure}[tbh!]
    \centering
    \includegraphics[width=\linewidth]{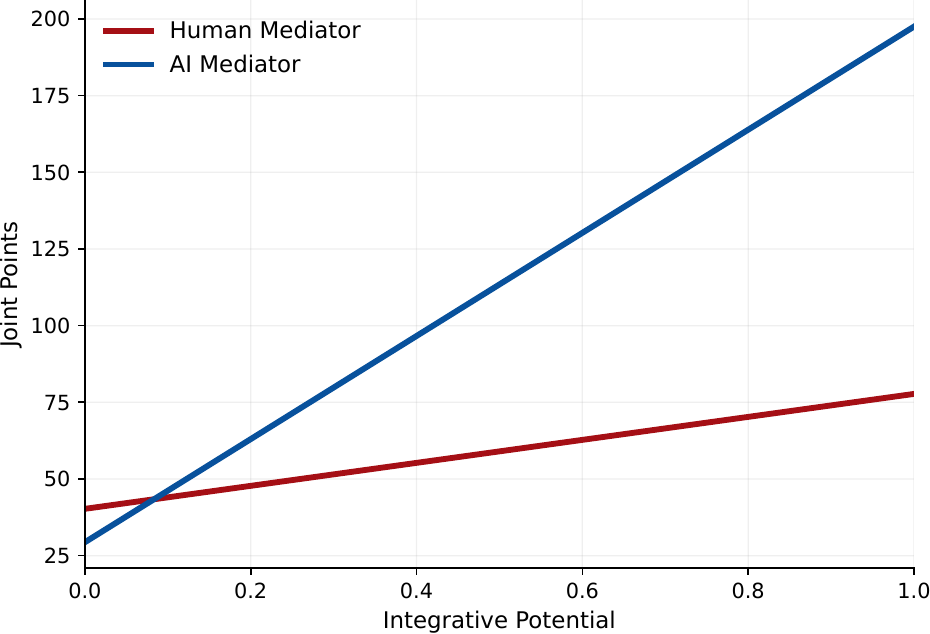}
    \caption{This graph depicts how an interaction between mediator type (human versus AI) and integrative potential affects joint points achieved by disputants. In disputes with high integrative potential, the AI mediator  ``grows the pie.'' }
    \label{fig:condition_ip_interaction_panels}
\end{figure}

\subsubsection{Proliferation of Positive Emotions}
Now, we move to examine whether mediation can induce more positive emotions. We combine \textit{compassion}, and \textit{joy} into a single measure, and repeat the RM ANOVA, as done previously. Again, we find a significant main effect of time ($F(1,154)=9.05$, $p<.01$), suggesting mediation aids in promoting positive emotion -- before ($M=0.12$, $SD=0.15$) versus after ($M=0.15$, $SD=0.18$). We find a near trend ($F(2,154)=2.24$, $p=.11$) of an interaction between Mediation Type and Time. By t-test, the AI mediator significantly ($t=-2.61$, $p=.01$) increases positive emotion going from before ($M=0.09$, $SD=0.13$) to after ($M=0.14$, $SD=0.20$); whereas, the other two conditions do not reach significance. Thus, to answer \textbf{RQ1}, regarding whether AI mediators can regulate emotion in contentious disputes, we find evidence they can. 

\subsection{AI Mediators Expand the Pie}
We present evidence AI mediators, relative to human ones, can expand the pie when there exists high integrative potential. We further analyze the types of messages mediators send, showing the AI sends more messages suggesting trade-offs.

\subsubsection{IP Interacts with Joint Points}\label{sec:jointpoitns}
We first test, via moderated regression, whether there exists an interaction between integrative potential and condition (only considering AI and human-mediated cases) on the points scored -- i.e., the value of the agreement to each side. 
We consider seller points, buyer points, and joint points as dependent variables; we consider condition (AI versus human mediator) and integrative potential (IP) as independent variables. We z-score points, and IP before the regressions.  
With respect to seller points, we find a significant interaction between condition and integrative potential ($\beta=0.39$, $p=.04$), where the AI mediator leads to more points in cases of high IP, relative to the human mediator. Moving to buyer points ($\beta=0.27$, $p=.16$), we do not see a significant interaction. However, joint points yields a marginally significant interaction ($\beta=0.37$, $p=.056$). 
Figure~\ref{fig:condition_ip_interaction_panels} illustrates the interaction for joint points -- the plots for buyer and seller points look similar. These results indicate AI mediators outperform human ones at realizing joint potential -- i.e., ``growing the pie.'' Thus, with respect to \textbf{RQ2}, asking whether LLMs can effectively facilitate trade-offs in disputes with IP, we find supportive evidence.

\subsubsection{AI Mediators Suggest More Trade-offs}
\begin{figure}
    \centering
    \includegraphics[width=.7\linewidth]{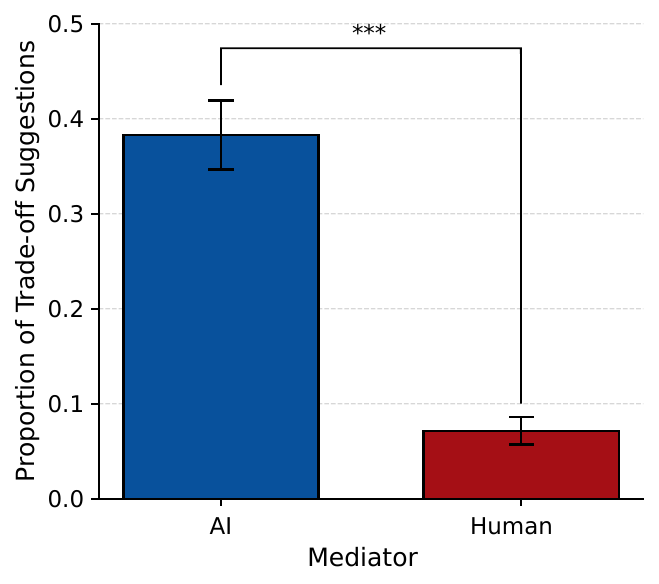}
    \caption{AI mediators suggest trade-offs significantly more than humans. }
    \label{fig:tradeoff}
\end{figure}

While we find LLMs can help disputants discover joint gains, we have yet to analyze the types of messages mediators send. 
As discussed in Section~\ref{sec:coding}, we create a dependent variable using the LLM annotations -- specifically, we calculate the proportion of places the mediator could have intervened (i.e., after each utterance), where they did intervene and suggested a trade-off. 
We run a t-test to compare between the AI and human mediators, finding a significant effect ($t=8.00$, $p<.001$), where the AI mediators send more ($M=0.38$, $SD=0.27$) trade-off suggestions compared to the human mediators ($M=0.07$, $SD=0.10$). 
Figure~\ref{fig:tradeoff} illustrates this difference. 
Thus, we find evidence that LLM compared to human mediators use more language supporting trade-offs, which may act as a mechanism to induce the increased joint gains found in Section~\ref{sec:jointpoitns} (\textbf{RQ2}); however, confirming this requires future studies. 

While we focus primarily on trade-off suggestions, Table~\ref{tab:negotiact} presents results for the other codes in the scheme, where at least one side's mean lands above $0.1$.
Generally, the LLM mediator sends more messages suggesting action, offer modification, procedure, trade-off, and time considerations; further, the LLM tends more to ask for substantiation, positional information, and clarification; lastly, the LLM engages in more active listening, on average. 
Each message might fit multiple codes -- on average, the AI's messages fit $M=3.12$ codes, while the humans' match with $M=1.07$ categories. Thus, the AI's mediation messages serve multiple functions.

\section{Discussion}
These results suggest AI mediators promote an interest-focused strategy, which prior work shows moves people off of angry entrenched positions and towards problem solving \cite{lytle1999strategic}.
In Section~\ref{sec:emotion}, we first demonstrate AI mediators significantly reduce negative emotions in disputes, while the human mediator does not; we also find the AI mediator increases positive emotion, while the human mediator and counterfactual condition do not -- prior work indicates negative emotion can derail potential resolution \cite{pruitt2007conflict}.
We would, then, expect to also see more joint gains in AI-mediated disputes compared to human ones -- in Section~\ref{sec:jointpoitns} we see a marginally significant interaction where integrative potential (IP) and mediator type interact, such that the AI leads disputants to discover more joint gains when IP is high. 
While we do not possess a causal mechanism showing regulating emotion creates joint gains, these results provide evidence for that, which aligns with the literature\cite{carnevale1986influence,isen2015relationship}. 
Additionally, we show the AI mediator works to suggest trade-offs significantly more often than the human, potentially an explanatory mechanism for realization of joint gains \cite{froman1970compromise, brett2016negotiation} -- this also implies LLM mediators may natively fit more into the camp of \textit{evaluative} mediation \cite{zumeta2000styles}, a paradigm where the mediator suggests solutions, rather than simply facilitating communication. 


\section{Conclusion \& Future Work}
This work demonstrates the potential of AI mediators to effectively regulate emotion in heated disputes, which could lead to more collaborative behavior and joint gains.
However, a primary limitation of this work stems from using \texttt{GPT4o} in the buyer's role, mirroring the approach by Hale et al. \cite{hale2025kodis}, which yields two concerns -- 1) whether other language models better model human disputants, and 2) to what extent results generalize from human-AI disputes to dyadic human ones. These concerns require future studies to settle. Further, we use novice crowd-sourced mediators, rather than trained professionals -- while certainly worthwhile to compare against novices, usage in high-stakes professional settings would require evaluation against trained mediators. Additionally, our work only considers text interactions; future studies might examine face-to-face disputes (over video or in-person), where disputants have access to visual and auditory modalities. 
Lastly, while our results focus on value creation and emotion regulation, we could not show AI mediators significantly impacted impasse rates -- thus, this remains an area of future exploration.

\section{Ethical Impact Statement}
There exist several ethical considerations with respect to using AI in dispute mediation. Firstly, as prior work points out, self-determination and voluntary participation remain quintessential aspects of the mediation process -- thus, an AI mediator must not infringe on those principles, harming one's self-determination or imposing a solution rather than facilitating a resolution\cite{izumi2010implicit,nolan2013judicial}.
Further, LLMs train on human text, which contains many of the same biases found in humans. Consequently, there exists much research on the topic of bias in AI\cite{gao2025measuring, wan2023kelly, konggender}. In mediation especially this becomes an issue as it would preclude the mediator from being a neutral facilitator. 

Secondly, in addition to using LLMs as mediators, we use them to annotate emotion and dialogue acts. However, the issue of bias once again arises, where LLMs may take an overly western perspective when generating output\cite{mihalcea2025ai}. While, in our case with American-based participants this issue may not seem overly severe, if we move to inter-cultural disputes we may inadvertently induce worse outcomes, or generate annotations which lead to incorrect inferences. Thus, one should take extra care when deploying LLM mediators in multi-cultural scenarios, and with the inferences derived from LLM-generated annotations.  

\section*{Acknowledgments}
 This work is supported by the U.S. Government including the Air Force Office of Scientific Research (grant FA9550-23-1-0320). The views and conclusions contained in this document are those of the authors and should not be interpreted as representing the official policies, either expressed or implied, of the U.S. Government. The U.S. Government is authorized to reproduce and distribute reprints for Government purposes notwithstanding any copyright notation herein. 
\bibliographystyle{IEEEtran}
\bibliography{sample}
\end{document}